\documentclass[12pt]{article}
\begin{document}
\title{A Note on a Particle-Antiparticle Interaction}
\author{B.G. Sidharth\\
International Institute for Applicable Mathematics \& Information Sciences\\
Hyderabad (India) \& Udine (Italy)\\
B.M. Birla Science Centre, Adarsh Nagar, Hyderabad - 500 063
(India)}
\date{}
\maketitle
\begin{abstract}
We develop an iso spin like formulation with particles and their
anti particle counterparts. This leads to a new shortlived
interaction between them, valid at very high energies and mediated
by massive particles. We point out that evidence for this is already
suggested by the very recent observations by the CDF team at Fermi
Lab.
\end{abstract}
It is well known that the Dirac equation \cite{greiner,bd} is
\begin{equation}
(\gamma^\mu p_\mu - m) \psi = 0\label{e1}
\end{equation}
Here $\gamma^\mu$ are $4 \times 4$ matrices obeying the Clifford
algebra and $\psi$ is a 4 component wave function (spinor). $\psi$
can be written as,
\begin{equation}
\psi = \left(\begin{array}{ll} \phi \\
\chi\end{array}\right)\label{e2}
\end{equation}
where $\phi$ and $\chi$ are 2-component spinors, $\phi$ being the
"large" or positive energy component of $\psi$ and $\chi$ is the
"small" or negative energy component which is such that
\begin{equation}
\chi \sim (\frac{v}{c})^2 \phi\label{e3}
\end{equation}
It is also known that this picture gets reversed at high energies
where $v \to c$ (Cf.refs.\cite{greiner,bd}).\\
We observe that (\ref{e1}) can be written as:
$$\imath \hbar (\partial \phi / \partial t) = c \tau \cdot (p -
e/cA) \chi + (mc^2+e\phi)\phi ,$$
\begin{equation}
\imath \hbar (\partial \chi / \partial t) = c \tau \cdot (p - e/cA)
\phi + (-mc^2+e\phi)\chi.\label{3.29}
\end{equation}
We can see from (\ref{3.29}) that (in the absence of electromagnetic
subsection),
\begin{equation}
t \to -t, \quad \phi \to -\chi\label{e5}
\end{equation}
Let us now consider intervals near the Compton scale, where as we
know $v \to c$, and $\chi$ no longer is the "small" component.\\
At the Compton scale we have the phenomenon of Zitterbewegung or
rapid unphysical oscillation. It has been pointed out that in this
case, \cite{uof} that time can be modelled by a double Weiner
process and can be described as follows
\begin{equation}
\frac{d_+}{dt} x (t) = {\bf b_+} \, , \, \frac{d_-}{dt} x(t) = {\bf
b_-}\label{2ex1}
\end{equation}
where for simplicity we consider in the one dimensional case. This
equation (\ref{2ex1}) expresses the fact that the right derivative
with respect to time is not necessarily equal to the left
derivative. It is well known that (\ref{2ex1}) leads to the Fokker
Planck equations \cite{tduniv,ness}
$$
\partial \rho / \partial t + div (\rho {\bf b_+}) = V \Delta \rho
,$$
\begin{equation}
\partial \rho / \partial t + div (\rho {\bf b_-}) = - U \Delta
\rho\label{2ex2}
\end{equation}
defining
\begin{equation}
V = \frac{{\bf b_+ + b_-}}{2} \quad ; U = \frac{{\bf b_+ - b_-}}{2}
\label{2ex3}
\end{equation}
We get on addition and subtraction of the equations in (\ref{2ex2})
the equations
\begin{equation}
\partial \rho / \partial t + div (\rho V) = 0\label{2ex4}
\end{equation}
\begin{equation}
U = \nu \nabla ln\rho\label{2ex5}
\end{equation}
It must be mentioned that $V$ and $U$ are the statistical averages
of the respective velocities and their differences. We can then
introduce the definitions
\begin{equation}
V = 2 \nu \nabla S\label{2ex6}
\end{equation}
\begin{equation}
V - \imath U = -2 \imath \nu \nabla (l n \psi)\label{2ex7}
\end{equation}
We refer the reader to Smolin \cite{smolin} for further details. We
now observe that the complex velocity in (\ref{2ex7}) can be
described in terms of a positive or uni directional time $t$ only,
but with a complex coordinate
\begin{equation}
x \to x + \imath x'\label{2De9d}
\end{equation}
To see this let us rewrite (\ref{2ex3}) as
\begin{equation}
\frac{dX_r}{dt} = V, \quad \frac{dX_\imath}{dt} = U,\label{2De10d}
\end{equation}
where we have introduced a complex coordinate $X$ with real and
imaginary parts $X_r$ and $X_\imath$, while at the same time using
derivatives with respect
to time as in conventional theory.\\
From (\ref{2ex3}) and (\ref{2De10d}) it follows that
\begin{equation}
W = \frac{d}{dt} (X_r - \imath X_\imath )\label{2De11d}
\end{equation}
This shows that we can use derivatives with respect to the usual
time derivative with the complex space coordinates (\ref{2De9d}) (Cf.ref.\cite{bgsfpl162003}.\\
Generalizing (\ref{2De9d}), to three dimensions, we end up with not
three but four dimensions,
$$(1, \imath) \to (I, \tau),$$
where $I$ is the unit $2 \times 2$ matrix and $\tau$s are the Pauli
matrices. We get the special relativistic \index{Lorentz}Lorentz
invariant metric at the same time.\\
That is,\\
\begin{equation}
x + \imath y \to Ix_1 + \imath x_2 + jx_3 + kx_4,\label{Aa}
\end{equation}
where $(\imath ,j,k)$ momentarily represent the \index{Pauli}Pauli
matrices; and, further,
\begin{equation}
x^2_1 + x^2_2 + x^2_3 - x^2_4\label{B}
\end{equation}
is invariant, thus establishing a one to one correspondence between
(\ref{Aa}) and Minkowski 4 vectors as shown by (\ref{B}).\\
In this description we would have from (\ref{Aa}), returning to the
usual notation,
\begin{equation}
[x^\imath \tau^\imath , x^j \tau^j] \propto \epsilon_{\imath jk}
\tau^k \ne 0\label{y}
\end{equation}
(No summation over $\imath$ or $j$) Alternatively, absorbing the $x^\imath$ and
$\tau^\imath$ into each other, (\ref{y}) can be written as
\begin{equation}
[x^\imath , x^j] = \beta \epsilon_{\imath jk} \tau^k\label{xa}
\end{equation}
Equation (\ref{y}) and (\ref{xa}) show that the coordinates no
longer follow a commutative geometry. It is quite remarkable that
the noncommutative geometry (\ref{y}) has been studied by the author
in some detail (Cf.\cite{tduniv}), though from the point of view of
Snyder's minimum fundamental length, which he introduced to overcome
divergence difficulties in Quantum Field Theory. Indeed we are
essentially in the same situation, because for our positive energy
description of the universe, there is the minimum Compton wavelength
cut off for a meaningful description as is well known
\cite{bgsextn,schweber,newtonwigner}. Following Feshbach and Villars
(loc.cit) we consider (\ref{e2}) to describe  particles and anti-particles : specifically a particle
anti-particle pair depending on the upper or lower component predominating.\\
Proceeding further we could invoke the $SU (2)$ and consider the
gauge transformation \cite{taylor}
\begin{equation}
\psi (x) \to exp [\frac{1}{2} \imath g \tau \cdot \omega (x)] \psi
(x).\label{4.2}
\end{equation}
This is known to lead to a gauge covariant derivative
\begin{equation}
D_\lambda \equiv \partial_\lambda - \frac{1}{2} \imath g \tau \cdot
\bar{W}_\lambda,\label{4.3a}
\end{equation}
We are thus lead to vector Bosons $\bar{W}_\lambda$ and an
interaction like the weak interaction, described by
\begin{equation}
\bar{W}_\lambda \to \bar{W}_\lambda + \partial_\lambda \omega - g
\omega \Lambda \bar{W}_\lambda.\label{4.4}
\end{equation}
However, we are this time dealing, not with iso spin, but between
positive and negative energy states as in (\ref{3.29}) that is
particles and antiparticles. Also we must bear in mind that this new
non-electroweak force between particles and anti particles would
be short lived as we are at the Compton scale \cite{report}.\\
These considerations are also valid for the Klein-Gordon equation
because of the two component formulation developed by Feshbach and
Villars \cite{feshbach,uheb}. There too, we get equations like
(\ref{3.29}) except that $\phi$ and $\chi$ are in this case scalar
function. We would like to re-emphasize that our usual description
in terms of positive energy solutions is valid above the Compton
scale (Cf.refs.\cite{greiner,bd}). To put it another way, equation
(\ref{e2}) describes a new spinor in a "superspin" space.\\
Thus we are lead to a new short lived interaction (as we are near
the Compton scale), mediated by vector Bosons $\bar{W}$.\\
With regard to the $\bar{W}$ acquiring mass, apart from the usual
approach, we can note the following. Equation (\ref{y}) underlines
the non-commutativity of spacetime, and under these circumstances it
has been argued that there is a break in symmetry that leads to a
mass being acquired exactly as with the Higgs mechanism
\cite{bgsijmpe,tduniv}.\\
Let us see this in greater detail. The Gauge Bosons would be
\index{mass}massless
and hence the need for a \index{symmetry breaking}symmetry breaking, \index{mass}mass generating mechanism.\\
The well known remedy for the above situation has been to consider,
in analogy with \index{superconductivity}superconductivity theory,
an extra phase of a self coherent system (Cf.ref.\cite{moriyasu} for
a simple and elegant treatment and also refs. \cite{jacob} and
\cite{taylor}). Thus instead of the \index{gauge field}gauge field
$A_\mu$, we consider a new phase adjusted \index{gauge field}gauge
field after the \index{symmetry}symmetry is broken
\begin{equation}
\bar{W}_\mu = A_\mu - \frac{1}{q} \partial_\mu \phi\label{Eex4}
\end{equation}
The field $\bar{W}_\mu$ now generates the \index{mass}mass in a self
consistent manner via a Higgs mechanism. Infact the kinetic energy
term
\begin{equation}
\frac{1}{2} |D_\mu \phi |^2\quad ,\label{Eex5}
\end{equation}
where $D_\mu$ in (\ref{Eex5}) denotes the gauge derivative, now
becomes
\begin{equation}
|D_\mu \phi_0 |^2 = q^2|\bar{W}_\mu |^2 |\phi_0 |^2 \, ,\label{Eex6}
\end{equation}
Equation (\ref{Eex6}) gives the \index{mass}mass in terms of the ground state $\phi_0$.\\
The whole point is as follows: The \index{symmetry breaking}symmetry
breaking of the \index{gauge field}gauge field manifests itself only
at short length scales signifying the fact that the field is
mediated by particles with large \index{mass}mass. Further the
internal \index{symmetry}symmetry space of the \index{gauge
field}gauge field is broken by an external constraint: the wave
function has an intrinsic relative phase factor which is a different
function of spacetime coordinates compared to the phase change
necessitated by the minimum coupling requirement for a free particle
with the gauge potential. This cannot be achieved for an ordinary
point like particle, but a new type of a physical system, like the
self coherent system of \index{superconductivity}superconductivity
theory now interacts with the \index{gauge field}gauge field. The
second or extra term in (\ref{Eex4}) is effectively an external
field, though (\ref{Eex6}) manifests itself only in a relatively
small spatial interval. The $\phi$ of the Higgs field in
(\ref{Eex4}), in analogy with the phase function of  \index{Cooper
pairs}Cooper pairs of
\index{superconductivity}superconductivity theory comes with a \index{Landau-Ginzburg}Landau-Ginzburg potential $V(\phi)$.\\
Let us now consider in the \index{gauge field}gauge field
transformation, an additional phase term, $f(x)$, this being a
scalar. In the usual theory such a term can always be gauged away in
the \index{U(1)}U(1) \index{electromagnetic}electromagnetic group.
However we now consider the new situation of a
\index{noncommutative}noncommutative geometry referred to above,
\begin{equation}
\left[dx^\mu , dx^\nu \right] = \Theta^{\mu \nu} \beta , \beta \sim
0 (l^2)\label{Eex7}
\end{equation}
where $l$ denotes the minimum \index{spacetime}spacetime cut off.
Equation (\ref{Eex7}) is infact \index{Lorentz}Lorentz covariant.
Then the $f$ phase factor gives a contribution to the second order
in coordinate differentials,
$$\frac{1}{2} \left[\partial_\mu B_\nu - \partial_\nu B_\mu \right] \left[dx^\mu , dx^\nu \right]$$
\begin{equation}
+ \frac{1}{2} \left[\partial_\mu B_\nu + \partial_\nu B_\mu \right]
\left[dx^\mu dx^\nu + dx^\nu dx^\mu \right]\label{Eex8}
\end{equation}
where $B_\mu \equiv \partial_\mu f$.\\
As can be seen from (\ref{Eex8}) and (\ref{Eex7}), the new
contribution is in the term which contains the commutator of the
coordinate differentials, and not in the symmetric second term.
Effectively, remembering that $B_\mu$ arises from the scalar phase
factor, and not from the non-Abelian gauge field,
$A_\mu$ is replaced by
\begin{equation}
A_\mu \to A_\mu + B_\mu = A_\mu + \partial_\mu f\label{Eex9}
\end{equation}
Comparing (\ref{Eex9}) with (\ref{Eex4}) we can immediately see that
the effect of noncommutativity is precisely that of providing a new
\index{symmetry breaking}symmetry breaking term to the gauge
field, instead of the $\phi$ term, (Cf.refs. \cite{cr39,ijmpe}) a term not belonging to the gauge field itself.\\
On the other hand if we neglect in (\ref{Eex7}) terms $\sim l^2$,
then there is no extra contribution coming from (\ref{Eex8}) or
(\ref{Eex9}), so that we are in the usual non-Abelian gauge field theory, requiring a broken
symmetry to obtain an equation like (\ref{Eex9}).
This is not surprising because if we neglect the term $\sim l^2$ in
(\ref{Eex7}) then we are back with the usual commutative theory and
the usual Quantum Mechanics.\\
It is quite remarkable that after the new interaction with the
$\bar{W}$ particles was proposed, the CDF team in Fermi Lab
announced a new force and particle in proton anti-proton
interactions that matches the above \cite{altonen}. The CDF rules
out the Higgs Boson because the decays are much too rapid for this
to be a Higgs Boson. The experimental result has been checked to a
$3 \sigma$ plus level of confidence, that is there is only a one in
thousand chance for it to be wrong.Subsequently this was pushed up to nearly five sigma the desired level though the Dzero could not show up this finding   If however the above theory and
experiment are talking about the same thing then surely the
confidence level increases. Further experimental results are
awaited. Other possible explanations included techni-colour.

\end{document}